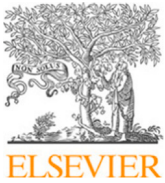
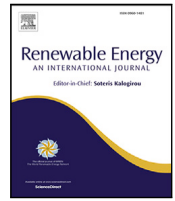

# Low-cost wind turbine aeroacoustic predictions using actuator lines

Laura Botero-Bolívar [a,*], Oscar A. Marino [b], Cornelis H. Venner [a], Leandro D. de Santana [d], Esteban Ferrer [b,c]

[a] *University of Twente, Drienerlolaan 5, Enschede, 7522 NB, Netherlands*
[b] *ETSIAE-UPM - School of Aeronautics, Universidad Politécnica de Madrid, Plaza Cardenal Cisneros 3, Madrid, E-28040, Spain*
[c] *Center for Computational Simulation - Universidad Politécnica de Madrid, Campus de Montegancedo, Boadilla del Monte, Av. de Montepríncipe, Madrid, 28660, Spain*
[d] *German-Dutch Wind Tunnels, Netherlands*

A B S T R A C T

Aerodynamic noise is a limitation for further exploitation of wind energy resources. As this type of noise is caused by the interaction of turbulent flow with the airframe, a detailed resolution of the flow is necessary to obtain an accurate prediction of the far-field noise. Computational fluid dynamic (CFD) solvers simulate the flow field but only at a high computational cost, which is much increased when the acoustic field is resolved. Therefore, wind turbine noise predictions using numerical approaches remain a challenge. This paper presents a methodology that couples (relatively fast) wind turbine CFD simulations using actuator lines with a fast turn-around noise prediction method. The flow field is simulated using actuator lines and large eddy simulations. The noise prediction method is based on the Amiet–Schlinker's theory for rotatory noise sources, considering leading- and trailing-edge noise as unique noise sources. A 2D panel code (XFOIL) calculates the sectional forces and boundary layer quantities. The resulting methodology for the noise prediction method is of high fidelity since the wind turbine geometry is accounted for in both flow and acoustics predictions. Results are compared with field measurements of a full-scale wind turbine for two operational conditions, validating the results of this research.

## 1. Introduction

The aerodynamic design of wind turbines has reached an almost optimal point in the last years due to the high investment in aerodynamic optimization and advances in manufacturing techniques and materials. Therefore, the noise produced by wind turbines is now becoming a competitive factor for the wind energy industry. Predicting wind turbine noise under real operational and atmospheric conditions as accurately as possible is paramount for designing more silent wind turbines and achieving the imposed noise limits [1]. Furthermore, fast turn-around methods are crucial to incorporate noise calculations during the design and optimization process of wind turbines and to assess the noise in real-time while the wind turbines are operating. This would optimize the harvest of the wind resources, minimizing the impact on the quality of life of the surrounding communities and wildlife.

The main noise source of typical horizontal axis modern wind turbines is aerodynamic noise, which is caused by the interaction of the flow with the wind turbine blades. Trailing-edge noise has been recognized as the dominant noise source of wind turbines [2]. It is caused by the turbulent eddies inside the boundary layer convecting past the blade trailing edge. Therefore, it is expected to appear in every operational and atmospheric condition. Significant efforts have been invested in reducing this type of noise using passive methods such as serrations [3] and consequently, other noise sources have become increasingly important. Nowadays, leading-edge noise, generated by the inflow turbulence impinging the blades, is an additional and relevant source of noise of wind turbines [4]. Leading-edge noise depends on the inflow turbulence characteristics of the atmospheric flow, and therefore its contribution to the total noise is not straightforwardly obtained. Thus, to properly predict wind turbine noise, a good resolution of the flow field around the wind turbine to obtain aerodynamic loadings and boundary layer data is fundamental to predicting trailing-edge noise, along with considering the inflow turbulence characteristics to predict leading-edge noise. It is this variety of scales, ranging from atmospheric turbulence to boundary layer flow scales, that makes acoustic simulations a very costly task.

There are several empirical methods to predict the overall noise level produced by a wind turbine based on the rated power and diameter, which are commonly used in noise regulations [5]. These

\* Corresponding author.
*E-mail address:* l.boterobolivar@utwente.nl (L. Botero-Bolívar).

**Nomenclature**

| Symbol | Description |
|---|---|
| $A_{\text{atm}}$ | atmospheric absorption [dB] |
| $AR$ | aspect ratio [–] |
| $B$ | number of blades [–] |
| $c$ | segment chord [m] |
| $c_0$ | speed of sound [m/s] |
| $C_d$ | drag coefficient [–] |
| $C_f$ | friction coefficient close to trailing edge [–] |
| $C_l$ | lift coefficient [–] |
| $C_p$ | power coefficient [–] |
| $C_T$ | thrust coefficient [–] |
| $\hat{CO}$ | unitary vector between convected noise source and observer location in the wind turbine frame [–] |
| $d$ | segment span for noise prediction [m] |
| $d_n$ | distance from the source to the segment center [m] |
| $d_m$ | distance from the mesh point to the blade axis [m] |
| $\vec{d_r}$ | translator vector to the mid of the segment [m] |
| $f$ | frequency [Hz] |
| $f_c$ | central frequency of one-third octave band [Hz] |
| $F_D$ | drag force [N] |
| $F_L$ | lift force [N] |
| $F_{x,y,z}$ | force in the x, y, and z-directions [N] |
| $H$ | boundary layer shape factor [–] |
| $l_y$ | correlation length in the spanwise direction of the segment [m] |
| $L$ | turbulent integral length scale [m] |
| $L_p$ | A-weighted far-field noise PSD [dBA] |
| $\vec{M_s}$ | segment Mach number [–] |
| $\vec{M_f}$ | flow Mach number [–] |
| $M_t$ | tangential Mach number of the segment [–] |
| $M_y, M_z$ | Rotational Matrixes [–] |
| $n$ | number of segments for the blade discretization [–] |
| $n_\psi$ | number of division of the rotor plane [–] |
| $P$ | pressure at the noise source [Pa] |
| $P_{\text{ref}}$ | reference pressure of the environment [Pa] |
| $r$ | radial position [m] |
| $\vec{R}$ | vector between observer and emission point [m] |
| $R_a$ | radial location of the mid of each segment [m] |
| $R_z$ | component of $\vec{R}$ in the $Z_{\text{WT}}$ direction [m] |
| $R0_{\text{WT}}$ | observer position on the fixed reference frame of the wind turbine [$(X_{\text{WT}}, Y_{\text{WT}}, Z_{\text{WT}})$] [m] |
| $R0_a$ | observer position on the Amiet's reference frame [$(x_a, y_a, z_a)$] [m] |
| $Re$ | local Reynolds number at mid of the segment [–] |
| $S_{pp|\text{seg}}$ | PSD of the total noise produced by each segment in a specific angular position [Pa$^2$/Hz] |
| $S_{pp|\text{LE}}$ | PSD of the leading-edge noise produced by each segment in a specific angular position [Pa$^2$/Hz] |
| $S_{pp|\text{blade}}$ | PSD of the total noise produced by the blade in a specific angular position [Pa$^2$/Hz] |
| $S_{pp|\text{WT}}$ | PSD of the total noise produced by the wind turbine over one rotation [Pa$^2$/Hz] |
| $S_{pp|\text{TE}}$ | PSD of the trailing-edge noise produced by each segment in a specific angular position [Pa$^2$/Hz] |
| $T$ | temperature at the noise source [k] |
| $T_{\text{ref}}$ | reference temperature of the environment [k] |
| $TSR$ | tip speed ratio [–] |
| $Tu_\infty$ | turbulence intensity (in the $Z_{\text{WT}}$ direction) [m/s] |
| $u_\tau$ | friction velocity close to trailing edge [m/s] |
| $u_\psi$ | azimuth component of local relative velocity [m/s] |
| $U_c$ | convection velocity close to the trailing-edge [m/s] |
| $U_{rel}$ | relative velocity [m/s] |
| $U_t$ | apparent tangential velocity of the segment [m/s] |
| $U_\infty$ | inflow velocity (in the $Z_{\text{WT}}$ direction) [m/s] |
| $\vec{x_c}$ | location of the convected noise source $\vec{x_e}$ [m] |
| $\vec{x_e}$ | location of the emitted noise source [m] |
| $x_{i,j,k}$ | mesh point [m] |
| $X_{\text{WT}}, Y_{\text{WT}}, Z_{\text{WT}}$ | axes in the fixed reference frame of the wind turbine, $Z_{\text{WT}}$ in the downwind direction, and $Y_{\text{WT}}$ in the vertical direction [m] |
| $x_a, y_a, z_a$ | axes in Amiet's reference frame, $x_a$ in chordwise direction, and $y_a$ in spanwise direction [m] |
| $\alpha$ | angle of attack [rad] |
| $\alpha_a$ | atmospheric attenuation [dB/m] |
| $\beta$ | local angle of incidence [rad] |
| $\Delta_{grid}$ | mesh size [m] |
| $\Delta_{x,y,z}$ | mesh size in the x, y, and z-directions [m] |
| $\omega$ | angular frequency [rad/s] |
| $\omega_e$ | emitted angular frequency [rad/s] |
| $\Omega$ | rotational speed [rad/s] |
| $\eta_\epsilon$ | Gaussian function to smear forces in AL model [–] |
| $\Psi$ | azimuthal angle [rad] |
| $\theta$ | pitch angle of the wind turbine [rad] |
| $\kappa$ | acoustic wavenumber (= $\omega/c_0$) [rad/m] |
| $\kappa_e$ | wavenumber of the largest eddies [rad/m] |
| $\kappa_x$ | aerodynamic wavenumber in the chordwise direction (= $\omega/U_t$) [rad/m] |
| $\phi$ | twist angle of each segment [rad] |
| $\theta$ | pitch angle of the blade [rad] |
| $u_\tau$ | friction velocity [–] |
| $\phi_{uu}$ | auto-power spectral density of velocity fluctuations in the $x_a$-direction [(m/s)$^2$/Hz] |
| $\phi_{ww}$ | auto-power spectral density of velocity fluctuations in the $x_a$-direction [(m/s)$^2$/Hz] |
| $\Phi_{pp}$ | auto-power spectral density of wall-pressure fluctuations close to the trailing edge [Pa$^2$/Hz] |
| $\Omega$ | rotational speed [rad/seg] |
| $\rho$ | air density [kg/m$^3$] |
| $\mathscr{L}$ | aeroacoustic transfer function for LE and TE noise prediction [–] |
| $\mathcal{R}$ | convective and viscous terms in the NS equations [–] |
| $\mathcal{S}(\mathbf{Q})$ | source term used by to include the actuator line forcing in NS equations [–] |
| $\mathbf{Q}$ | vector of conservative variables [–] |
| AL | actuator line |
| Avg. | average results from the AL simulations |
| BEMT | blade element momentum theory |
| inst. | instantaneous results from the AL simulations |
| LE | leading edge |
| NS | Navier–Stokes |
| O.C. | operational condition |
| PS | pressure side |
| PSD | power spectral density |
| SS | suction side |
| TE | trailing edge |
| OASPL | overall A-weighted sound pressure level |
| TUD | turbulence distortion |

| **Subscripts** | |
|---|---|
| a | Amiet's reference frame |
| b | blade reference frame |
| LE | leading edge |
| seg | segment |
| TE | trailing edge |
| WT | wind turbine inertial reference frame |

engineering methods are inaccurate when predicting noise in the frequency domain or when dealing with not nominal operational and atmospheric conditions, e.g., skewed blades, varying turbulence levels, etc. Very costly alternatives are available through computational aeroacoustic (CAA) models, where the near flow-field is simulated using computational fluid dynamics (CFD) approaches while the far-field noise propagation is computed using an acoustic analogy; such as Ffowcs–Williams and Hawkings (FWH) [6], which requires the hydrodynamic pressure fluctuations over a region as an input. The FWH acoustic analogy has been developed for rotating machinery either in the time domain [7,8] or in the frequency domain [9–11]. The main limitation of CAA methods is the high computational costs since the flow needs capture the evolution of turbulent eddies interacting with the blade surface to obtain the pressure fluctuations responsible of noise generation. Conducting Large Eddy Simulations (LES) with this level of detail remains impractically expensive. A cheaper alternative is to use the Reynolds Averaged Navier–Stokes (RANS) equations to model turbulence, as conducted by Tadamasa and Zangeneh [12]. This approach provides reasonable results for low frequencies but is difficult to generalize to more complex atmospheric or operational conditions. A midground approach is using strip-theory methods, where the wind turbine noise is calculated by dividing each blade into uncorrelated blade segments, and the noise of each segment is calculated using prediction methods for 2D airfoils. This method was proposed by Schlinker and Amiet [13] for helicopters and has recently been applied to fans [14], wind turbines [15–17], and open propellers [18]. Two main methods are available in the literature for predicting the trailing-edge noise of 2D airfoils. An empirical method proposed by Brooks et al. [19], known as the Brooks–Pope and Marcolini (BPM) model, is based on boundary layer and far-field noise measurements of several NACA 0012 airfoils with different chords. Although very useful for quick estimations, the main limitation of this method is that the airfoil geometry and operational conditions of wind turbines differ significantly from a NACA 0012 airfoil and the test conditions used by Brooks et al. [19]. Therefore, the prediction for general wind turbines may be inaccurate. A semi-analytical approach was proposed by Amiet [20], who used analytical solutions to calculate an aeroacoustic transfer function between the wall-pressure spectrum close to the trailing-edge and the far-field noise spectrum. The geometry of the airfoil and operational conditions are accounted for in the calculation of the wall-pressure spectrum model, which depends on boundary layer quantities. Similarly, Amiet [21] proposed a method for calculating leading-edge noise using as input the incoming turbulence spectrum.

Strip theory methods are low-cost methods for predicting wind turbine noise in the frequency domain, also considering geometric details for a variety of operating conditions. These methods alone cannot predict the flow field around the wind turbine, and need to be coupled to numerical simulations. Here, we propose to simulate the flow field around the wind turbine and blades using actuator lines (AL). AL is a computationally efficient approach for capturing the complex aerodynamic behavior of rotating blades by representing them as a collection of rotating actuator lines. AL models have been widely validated in the literature to properly capture the effect of wind turbine blades on the flow field [22–24]. In this method, each blade is modeled as a line, which is discretized in segments (following blade element momentum theory). Each blade segment provides a body force in the Navier–Stokes equations to mimic the effect of the rotating blades. AL models require 2D sectional data. This model has been coupled with the BPM model to calculate the trailing-edge noise of wind turbines, either directly [25], or coupled with linearized equations, such as the Acoustic Perturbation Equation (APE) [26], where the AL+BPM method is used as an input to the APE [27–29]. These ideas are the basis of our approach. Let us remind the reader that the fundamental problem of the BPM approach is that it does not take into account the geometry of the airfoil at each section and that it is based on experimental measurements only.

This research implements a wind turbine noise prediction method based on the AL model coupled with the Amiet–Schlinker theory for rotatory noise sources [13], considering the leading- and trailing-edge noise as the unique and independent noise source of a wind turbine. AL method is used to simulate the flow field around the turbine, rotor and blades and is completed with 2D XFOIL calculations when the boundary layer data is required to compute acoustics. An essential factor of this implementation is that the boundary layer parameters are obtained with 2D XFOIL simulations, which makes the prediction method consider details of the wind turbine geometry since the airfoil and angle of attack distribution along the blade is accounted for, and a fast turnaround method since Amiet's theory is very cheap computationally and XFOIL simulations are relatively quick compared to most of the numerical analyses. In the method proposed in this research, the wind turbine noise can be obtained considering a time-averaged blade loading or instantaneous one at each azimuthal location, which can account for non-axisymmetric or transient phenomena.

The remaining part of the paper is organized as follows. Section 2 introduces the AL model and the numerical method for the CFD simulations. Section 3 presents the theoretical approach of the wind turbine noise prediction method. Section 4 addresses the validation case, where convergence and sensitivity analyses are conducted, followed by a comparison of the predicted wind turbine noise with field measurements. Finally, Section 5 summarizes the main contributions and discussion of this work.

## 2. Actuator line simulations

Actuator Line [22] model the aerodynamic behavior of wind turbine blades without meshing the blade geometry. This method represents the blades as rotating lines, and introduces source terms into the flow equations that mimic the presence of the blades, allowing for a moderately accurate representation of blade aerodynamics and wake dynamics.

In the AL method, each blade is discretized as a line of multiple points spanning from the root to the tip of the blade. These lines are embedded within the computational domain, providing rotating forces. To incorporate the sectional blade forces, the AL employs blade element theory, which calculates the aerodynamic forces and moments on each line element based on its local angle of attack, airfoil characteristics, and inflow conditions. These forces are then distributed to the surrounding flow in the computational domain, influencing the flow velocity and pressure.

The forces are calculated by introducing tabulated lift and drag coefficients for different angles of attack and Reynolds numbers, typically obtained from a 2D numerical solution, e.g., XFOIL. The velocity of each node on the blade needs to be sampled to impose the forcing, but the force itself may influence this location. For this reason, various methods have been proposed to calculate the velocity (see [30]).

We follow a similar approach as Liu et al. [31] to compute the fluid kinematics around the blade sections. We start by considering the global inertial axes, $[X_{WT}, Y_{WT}, Z_{WT}]$, as shown in Fig. 1, where $X_{WT}$ represents the direction perpendicular to the flow velocity, $Y_{WT}$ is the local vertical axis, and $Z_{WT}$ is the direction of the incident flow

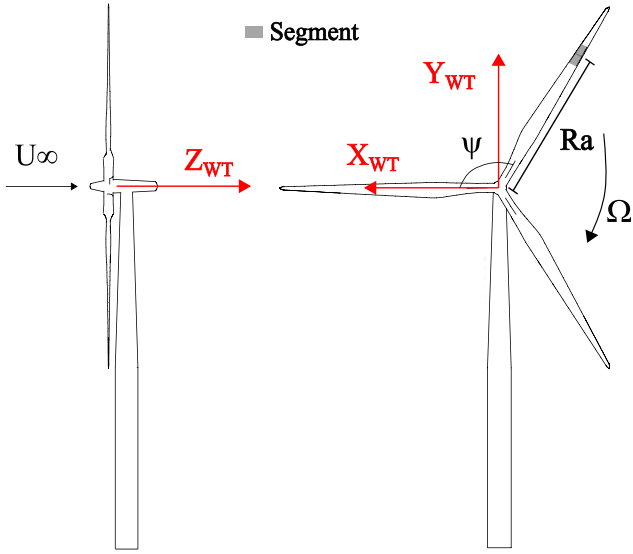

**Fig. 1.** Coordinate system of the wind turbine.

(perpendicular to the wind turbine rotor). These axes coincide with the CFD coordinate system.

The fluid velocity in this reference frame is be defined as $\vec{u} = [u_{x_{WT}}, u_{y_{WT}}, u_{z_{WT}}]$. The azimuthal component of the local velocity seen by a blade section is a combination of two components of the fluid velocity and the rotational speed, calculated as:

$$u_\Psi = \Omega R_a + u_{x_{WT}} \sin\Psi - u_{y_{WT}} \cos\Psi, \qquad (1)$$

and the correspondent angle of attack is defined as:

$$\alpha = \beta - \gamma; \quad \beta = \arctan\left(\frac{u_{z_{WT}}}{u_\Psi}\right); \qquad (2)$$

where $\beta$ is the local angle of incidence, and $\gamma = \phi + \theta$ is the angle due to the local twist of the airfoil ($\phi$) and pitch of the blade ($\theta$). All the velocities, angles, and forces can be seen in Fig. 2.

The relative velocity, $U_{rel} = \sqrt{u_{z_{WT}}^2 + u_\Psi^2}$, is obtained, and the forces exerted by the blade section to the fluid are projected onto the global axes:

$$\begin{aligned} F_{x_{WT}} &= (F_L \sin(\beta) - F_D \cos(\beta))\sin(\Psi), \\ F_{y_{WT}} &= -(F_L \sin(\beta) - F_D \cos(\beta))\cos(\Psi), \\ F_{z_{WT}} &= -(F_L \cos(\beta) + F_D \sin(\beta)); \end{aligned} \qquad (3)$$

where $F_L = 0.5\rho U_{rel}^2 S C_l$ and $F_D = 0.5\rho U_{rel}^2 S C_d$ are the lift and drag forces obtained with the lift and drag coefficients ($C_l$ and $C_d$ from XFOIL). These forces need to be distributed within a certain volume containing multiple grid cells to avoid simulation divergence. A Gaussian function [31,32] is typically used to smear the forces:

$$\eta_\epsilon = \frac{1}{\epsilon^3 \pi^{\frac{3}{2}}} e^{-\left(\frac{d}{\epsilon}\right)^2}; \qquad (4)$$

where $d_m$ is defined as the distance from the mesh point $x_{i,j,k}$ to the blade section ($x_b, y_b, z_b$), being both positions defined in the global axes,

$$d_m = \sqrt{(x_{i,j,k} - x_b)^2 + (y_{i,j,k} - y_b)^2 + (z_{i,j,k} - z_b)^2}. \qquad (5)$$

The parameter $\epsilon$ determines the width of the Gaussian projection. It is preferable for $\epsilon$ to be as small as possible to obtain a numerical simulation that closely resembles the initial model. However, it cannot be too small as one must avoid introducing singularities in the simulation. To this end, $\epsilon \sim 2\Delta_{grid}$, where $\Delta_{grid} = (\Delta_x \Delta_y \Delta_z)^{\frac{1}{3}}$, and $\Delta_{x,y,z}$ is the mesh size in the x, y, and z directions, as suggested by Martínez-Tossas et al. [33] and follow the general guideline $\epsilon \in [\Delta_{grid}, 4\Delta_{grid}]$.

Finally, the forces smeared by the Gaussian distribution are included as source terms in the spatially discretized Navier–Stokes equations, in their conservative form, as:

$$\frac{d\mathbf{Q}}{dt} = \mathcal{R}(\mathbf{Q}, \nabla\mathbf{Q}) + \mathcal{S}(\mathbf{Q}), \qquad (6)$$

where $\mathbf{Q}$ is the vector of conservative variables, $\mathbf{Q} = [\rho, \rho u_x, \rho u_y, \rho u_z, \rho e]^T$, $\mathcal{R}$ represents the convective and viscous terms and $\mathcal{S}(\mathbf{Q}) = \eta_\epsilon \mathbf{F}$ is the source term used to include the actuator line forcing.

The AL method is implemented in the high-order discontinuous Galerkin HORSES3D [34] to solve the compressible NS equations [35,36]. In HORSES3D, the physical domain is divided into different elements, each of them complemented with Gauss–Legendre nodes of 4th polynomial order (5th order accuracy). When the polynomial order of the basis is increased, the method shows spectral convergence (for smooth flows), leading to very accurate solutions.

## 3. Wind turbine noise prediction

The noise produced by the wind turbine is calculated using strip theory, where the blade is divided into *n* segments. Section 3.1 explains the methodology to split the blade. For each segment, leading- and trailing-edge noise is calculated as uncorrelated noise sources using Amiet's method for 2D airfoils. The noise is calculated for the effective or relative velocity ($U_{rel}$) of each segment obtained from the AL method, which accounts for the rotational velocity ($\Omega r$), the induced velocity at each segment location ($U_{ind}$), and the inflow velocity ($U_\infty$). Afterward, the total noise of the blade is calculated as the sum of the total noise of each segment. This procedure is done for each angular position $\Psi$.

The relative motion of the segment that induces a delay between the noise emission and observer locations is considered by a Doppler effect factor ($= \omega_e/\omega$). The total noise of the blade at each azimuth location is calculated by summing the noise produced by all the segments as:

$$S_{pp|\text{blade}}(\omega, \Psi) = \sum_1^n S_{pp|\text{seg}}(\omega_e, \Psi)(\omega_e/\omega)^2; \qquad (7)$$

where $S_{pp|\text{blade}}(\omega, \Psi)$ is the total noise of the blade as a function of the angular frequency ($\omega$) for each azimuthal angle ($\Psi$). Note that the flow is expected to be the same at each azimuth location, however, due to the directivity of the noise, the noise perceived by the observer is different at each azimuth location. $S_{pp|\text{seg}}(\omega_e)$ is the total noise of each segment calculated at the emission angular frequency ($\omega_e$), which is calculated as explained in Section 3.3. An exponent of 2 in the Doppler factor ($\omega_e/\omega$) in Eq. (7) is considered following the methodology of Sinayoko et al. [18]. The total noise of each segment is calculated as the sum of leading- and trailing-edge noise ($S_{pp|\text{seg}}(\omega_e) = S_{pp|\text{LE}}(\omega_e) + S_{pp|\text{TE}}(\omega_e)$).

The average noise produced by the wind turbine in one rotation ($S_{pp|\text{WT}}(\omega)$) is then calculated as:

$$S_{pp|\text{WT}}(\omega) = \frac{B}{2\pi} \int_0^{2\pi} S_{pp|\text{blade}}(\omega, \Psi) d\Psi; \qquad (8)$$

where $B$ is the number of blades.

The noise prediction of each segment is calculated using Amiet's theory for 2D airfoils. The location of the observer relative to each segment is transformed in the coordinates of Amiet's theory following the procedure described in Section 3.2.

### 3.1. Blade sections definition

The blade is divided into segments that are more refined close to the blade tip. An initial sinusoidal distribution of the location of the segments is proposed. The sinusoidal distribution is obtained by the horizontal coordinate of a point located in a semicircle of a diameter equal to the rotor radius (neglecting the inner part of the blades that consist of cylinders). The angle between the points in the radial axis was constant. After an iterative process to ensure that the aspect ratio ($AR$), defined as the span-chord ratio of the blade section, is larger than three, the radial position and the span- and chord-length distribution are obtained.

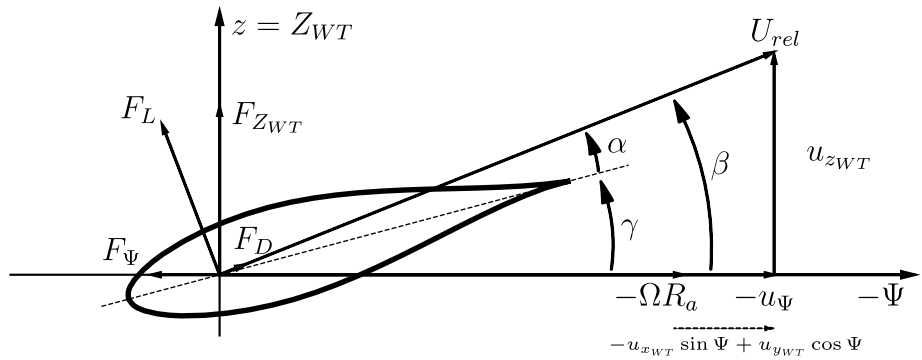

**Fig. 2.** Diagram of velocities and forces on the cross-sectional blade section.

For each segment, the geometric characteristics of the wind turbine, i.e., $c$, and $\phi$, and the blade loading, i.e., $Re$ and $\alpha$, are interpolated linearly from the wind turbine geometry information to obtain the values at the center of the section. As the airfoil might change along the segment, XFOIL simulations are conducted considering the airfoil at the extreme of the segment closer to the tip. With this approach ($AR \geq 3$), a maximum number of divisions for a given wind turbine blade radius and chord distribution is obtained.

### 3.2. Coordinate transformation

The coordinate system in the fixed reference frame of the wind turbine is $Z_{WT}$ located perpendicular to the rotor plane, positive in the downwind direction, and $Y_{WT}$ located in the vertical direction, positive upwards. $\Psi = 0$ is aligned with $X_{WT}$-axis. The origin of the coordinate system is the wind turbine hub; see Fig. 1.

To predict the noise of each segment, the observer location $\vec{R0}_{WT}$, given in the reference frame of the wind turbine, needs to be transformed to the Amiet's coordinate system ($\vec{R0}_a$). The first rotation is conducted in the $Z_{WT}$-axis to obtain $R0_{WT}$ in the blade reference frame $\vec{R0}_{blade}$ as: $\vec{R0}_{blade} = M_z \vec{R0}_{WT}$, where $M_z$ is:

$$M_z = \begin{bmatrix} \cos\Psi & \sin\Psi & 0 \\ -\sin\Psi & \cos\Psi & 0 \\ 0 & 0 & 1 \end{bmatrix}. \quad (9)$$

Later, a translation to the radial position of the segment ($R_a$) is conducted, i.e., $\vec{R0}_{seg} = \vec{R0}_{blade} + \vec{d}_r$, where $\vec{d}_r$ is defined as:

$$d_r = \begin{bmatrix} 0 \\ -R_a \\ 0 \end{bmatrix}; \quad (10)$$

$R_a$ is the radial location of the segment located at the center of the segment.

Finally, a rotation in the $y_{seg}$-axis is conducted to align the coordinate system with the segment chord. The rotational angle is the sum of the pitch angle of the blade ($\theta$) plus the twist angle of the segment ($\phi$). The final vector of the observer in Amiet's reference frame is $\vec{R0}_a = M_y \vec{R0}_{seg}$, where $M_y$ is:

$$M_y = \begin{bmatrix} \cos(\theta+\phi) & 0 & \sin(\theta+\phi) \\ 0 & 1 & 0 \\ -sin(\theta+\phi) & 0 & \cos(\theta+\phi) \end{bmatrix}. \quad (11)$$

In Amiet's reference frame, $x_a$ is located in the chordwise direction and $y_a$ in the spanwise direction.

### 3.3. Emission frequency

To calculate the Doppler factor, the methodology proposed by Sinayoko et al. [18] is adopted. They defined the Doppler factor as:

$$\frac{\omega_e}{\omega} = \frac{1 + (\vec{M}_f - \vec{M}_s) \cdot \hat{CO}}{1 + \vec{M}_f \cdot \hat{CO}}; \quad (12)$$

where $\vec{M}_f = [0, 0, U_\infty/c0]$ is the flow Mach number and $c0$ is the speed of sound, $\vec{M}_s$ is the segment Mach number ($\vec{M}_s = (\Omega R_a/c0)[-\sin\Psi, \cos\Psi, 0]$), and $\hat{CO}$ is the unitary vector between the convected noise source ($\vec{x}_c$) and the observer location in the frame of the wind turbine ($\vec{R0}_{WT}$), calculated as:

$$\hat{CO} = \frac{\vec{x}_c - \vec{R0}_{WT}}{|\vec{x}_c - \vec{R0}_{WT}|};$$

where $\vec{x}_c$ is the location of the convected noise source generated at $\vec{x}_e$, i.e., $\vec{x}_e$ is located at the middle of the blade segment ($\vec{x}_e = -R_a[-\sin\Psi, \cos\Psi, 0]$), whereas $\vec{x}_c$ accounts for the shift of the noise location because of the velocity, calculated as:

$$\vec{x}_c = \vec{x}_e + \vec{M}_f c_0 T_e; \quad (13)$$

where $c_0$ is the speed of sound and $T_e$ is the propagation time [37]:

$$c_0 T_e = \frac{-|M_f|R_z + \sqrt{|M_f|^2 R_z^2 + (1 - |M_f|^2)|R|^2}}{1 - |M_f|^2}; \quad (14)$$

where $\vec{R}$ is the vector between the observer and the emission point ($\vec{R} = \vec{R0}_{WT} - \vec{x}_e$), and $R_z$ is the component of $\vec{R}$ in the $Z_{WT}$ direction ($R_z = \vec{R} \cdot [0, 0, 1]$).

### 3.4. Leading-edge and trailing-edge noise prediction

Leading-edge (LE) and trailing-edge (TE) noise are predicted using Amiet's theory. The theory assumes a flat plate geometry with an infinitely small thickness, a stationary observer, and a uniform free-stream condition along the span. Amiet's theory calculates the far-field power spectral density of a flat plate of chord $c$ and span $d$. In LE and TE formulations, $k$ ($= \omega/c0$) is the acoustic wavenumber, and $k_x$ ($= \omega/U_t$) and $k_y$ are the chordwise and spanwise hydrodynamics wavenumbers, respectively. $k_y$ is assumed as 0 in the far-field approximation used for this case (valid for $AR \geq 3$ [38,39]). $\sigma^2$ ($= x_a^2 + (1 - M_t^2)(y_a^2 + z_a^2)$) is the flow corrected radial distance, where $M_t$ is the apparent Mach number at the center of the blade section, and $x_a$, $y_a$, and $z_a$ are the coordinates of the observer location in the Amiet's reference.

#### 3.4.1. Leading-edge noise
The power spectral density of the far-field LE noise at the midchord and midspan established by Amiet's theory [21] is:

$$S_{pp|LE} = \left(\frac{\omega z_a \rho c}{c_o \sigma^2}\right)^2 \pi U_t d \, |\mathcal{L}|^2 \, \Phi_{ww}; \quad (15)$$

where $\Phi_{ww}$ is the 2D velocity spectrum of the component perpendicular to the wall and $\mathscr{L}$ is the aeroacoustic transfer function for subcritical or supercritical gusts. $\mathscr{L}$ accounts for the scattering effect of the LE and back-scattering of the TE, implemented with the conclusions of Bresciani et al. [40, Eqs. A1–A4]. The coordinate reference system in Amiet's theory for LE is located at midchord and midspan. Therefore, an additional translation subtracting $c/2$ in the $x_a$-axis is conducted for the LE noise prediction.

The velocity spectrum can be calculated using the von Kármán method [41], the Liepmann method [42], and the turbulence distortion method [43]. The inputs for the three methods are the integral length scale and the turbulence intensity of the atmospheric inflow. Section 4.3.2 shows the prediction of the wind turbine noise using the different inflow turbulence models. The final comparison with the field measurements is conducted using the von Kármán spectrum model since it has been demonstrated that this model successfully describes atmospheric flows [44–46].

### 3.4.2. Trailing-edge noise

In addition to the assumptions discussed at the beginning of this section, for the TE noise formulation, the turbulence is assumed to be frozen at the TE discontinuity. Amiet's theory assumes that the origin of the coordinate system is at the TE location. The power spectral density of the TE noise using Amiet's theory [47] is:

$$S_{pp|\text{TE}} = \left(\frac{\omega z_a b}{4\pi c_o \sigma^2}\right)^2 d\, |\mathscr{L}|^2\, l_y \Phi_{\text{pp}}; \quad (16)$$

where $\mathscr{L}$ is the aeroacoustic transfer function that accounts for the subcritical and supercritical gusts and for the scattering and back-scattering propagation of trailing and leading-edges, modeled according to Roger and Moreau [39], $\Phi_{\text{pp}}$ is the wall-pressure spectrum close to the trailing edge, $l_y$ is the spanwise correlation length, and $b$ is the semichord. The spanwise correlation length is calculated as proposed by Corcos [48]:

$$l_y(\omega) = \frac{b_c U_c}{\omega}; \quad (17)$$

where $U_c$ is the convection velocity, assumed as $0.7U$, and $b_c$ is the Corcos' constant equal to 1.47 [49].

Several models are used for the calculation of the wall-pressure spectrum, i.e., TNO-Blake model [49], Kamruzzamann model [50], Lee model [51], Goody's model [52], and Amiet's model [47]. The inputs for all those models are the boundary layer characteristics close to the trailing edge, namely, boundary layer thickness ($\delta$), boundary layer displacement thickness ($\delta^*$), and friction velocity ($u_\tau$), which are calculated with XFOIL simulations.

The viscous panel code XFOIL simulations [53] gives the inputs to calculate the wall-pressure spectrum. The input parameters for XFOIL are the airfoil coordinates and the Reynolds number, based on the mean local chord of the section and the apparent tangential velocity, and the angle of attack, both obtained from the AL models. The location of the transition for XFOIL simulations is fixed at $x/c = 0.05$ (both for the pressure and suction sides) since the transition is expected close to the LE for the cases of full-scale wind turbines due to the high Reynolds number, inflow turbulence, and contamination and roughness of blade surface. Inflow turbulence is only considered in XFOIL through the N-critical number to switch the transition location. However, XFOIL is not able to incorporate other effect due to high turbulence and therefore, these are not taken into account in this research.

The boundary layer displacement thickness ($\delta^*$), momentum thickness ($\theta$), and skin friction coefficient ($C_f$) along the chord are obtained from XFOIL. The input needed to calculate the wall-pressure spectrum can be calculated based on that information. The boundary layer thickness ($\delta$) is calculated as [53]:

$$\delta = \theta \left(3.15 + \frac{1.72}{H-1}\right) + \delta^*; \quad (18)$$

**Table 1**  
Operational conditions of the wind turbine.

| Operational condition | $U_\infty$ [m/s] | $\Omega$ [rpm] | $\theta$ [deg] |
|---|---|---|---|
| O.C. 1 | 9.5 | 17 | 5 |
| O.C. 2 | 6.0 | 13 | 3 |

where $H = \delta^*/\theta$. The friction velocity ($u_\tau$) is calculated as:

$$u_\tau = \sqrt{0.5 U_t^2 C_f}. \quad (19)$$

The computational approach to predict wind turbine noise linked with XFOIL is openly available in [54].

The methodology proposed in this research is valid for cases where the boundary layer is attached. In cases of flow separation, a different approach should be considered. It has been demonstrated that Amiet's theory can be used for cases when separation flow occurs for 2D airfoils, e.g., at high angles of attack, if an appropriate wall-pressure spectrum model is used [55]. Bertagnolio et al. [56] proposed a semi-empirical wall-pressure spectrum model for separated boundary layers using as input the Reynolds number and the separation location. Cotté et al. [55] extended this model and coupled it with Amiet's theory to predict the separation noise of 2D airfoils, showing good agreement with experimental measurements. Recently, Suresh et al. [57] used Bertagnolio's method coupled with Amiet's theory to predict the separation noise of a small wind turbine. The separation location was obtained with numerical simulations. The approach used in this work may indicate where mild flow separation occurs based on the outputs from XFOIL, e.g., negative friction coefficient. However, to predict the location of flow separation (and predict separation noise), RANS/LES or other methods would be preferable. Finally, note that the noise prediction method discussed in this research can still be used if a wall-pressure spectrum model for separated boundary layers is used.

## 4. Test case and validation

### 4.1. Case definition

The test case is the Siemens SWT-2.3-93 wind turbine, a three-blade horizontal axis wind turbine located in the Høvsøre Wind Turbine Test Center in the northwest of Denmark. It has a nominal rated power of 2.3 MW. The rotor diameter is 93 m, and the hub height is 80 m.

Noise measurements at various operating conditions are reported in Leloudas [58]. The acoustic and met-mast measurements, operational curves, and wind turbine CAD are available in the database Christophe et al. [59]. We use the described methodology to compute the wind turbine noise for the two operational conditions summarized in Table 1. The inflow turbulence conditions at the hub height are calculated from atmospheric LES simulations reported in Kale [46] for one specific operational condition. Note that the same turbulence parameters are used for other operational conditions, assuming that the atmospheric turbulence does not vary significantly with the wind speed. The turbulence intensity and turbulence integral length scale at the hub height, needed to predict LE noise, are $Tu_\infty = 10.7\%$, and $L = 300$ m. The noise is measured at the ground at 100 m downwind of the wind turbine. Therefore, $X_{\text{WT}} = 0$ m, $Y_{\text{WT}} = 80$ m, and $Z_{\text{WT}} = 100$ m.

### 4.2. Aerodynamic simulations using actuator lines

The wind turbine simulations are performed using a Large Eddy Simulation turbulent model coupled with the actuator line. Simulations do not include the wind turbine tower. The AL considers the blades divided into 23 segments. The mesh used is a fully Cartesian grid in a prismatic domain of dimensions $[18R, 10R, 28R]$, being $R$ the rotor radius, with a constant size region near the turbine and stretching in the far field for all the directions except the vertical one, where it is performed at the top only. This results in a mesh consisting of approximately 99,600 hexahedral elements, resulting in 12.45 millions of degrees of freedom (DoF), per flow equation due to the 4th order polynomial approach. At the central refined region, there are approximately 29 DoF per blade radius.

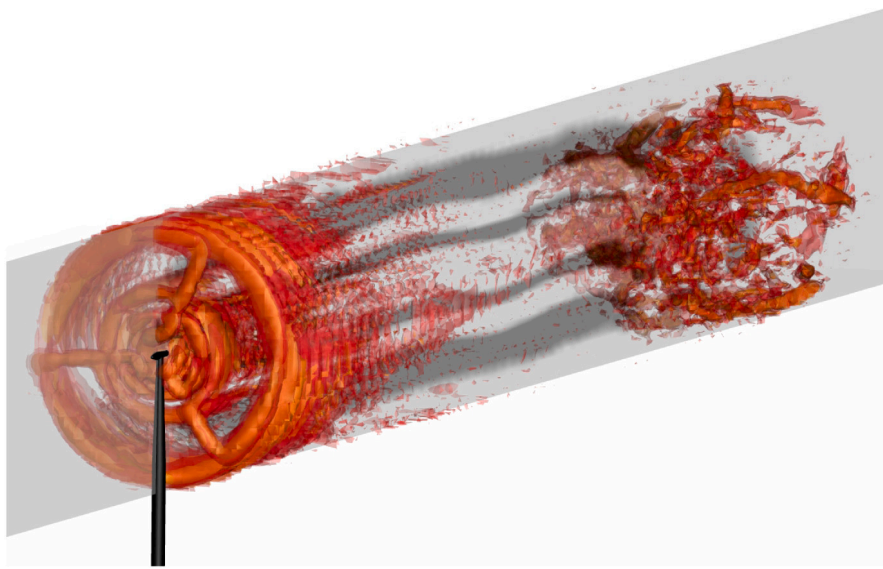

**Fig. 3.** Iso-contours of the Q-criteria and vertical slice of the magnitude of the velocity for the AL simulation for O.C. 1. The tower and nacelle are included for visualization purposes only.

AL-LES simulations are conducted parallelized in a cluster using a single node with 40 cores. The time for the AL simulations of this test case was about 260 CPU hours per wind turbine revolution (about 6.5 real hours).

In Fig. 3, the flow visualization of the AL simulation of O.C. 1 is presented using the Q-criteria. It can be seen that downstream the rotor plane up to two times the rotor diameter, the identity of the wake is maintained, clearly reflecting the rotational motion of the blades. Further downstream, the structures break down due to enhanced shear and mixing and tend to become more homogeneous. Fig. 3 also shows a vertical slice of the velocity, where the transition to turbulence of the wake is observed. The AL methodology implemented in the high-order solver captures the tip vortex and wake development with a considerable level of detail.

To verify the aerodynamic results of the AL simulations, we use blade element momentum theory (BEMT) simulations of the wind turbine under the same operational conditions to compare the noise inputs with AL simulations. The BEMT solutions are obtained using the OpenFAST code [16], which includes the Prandtl tip and root loss factor and the Pitt/Peters skewed-wake correction model. Results are presented in Fig. 4, where the time-averaged local apparent velocity and angle of attack along one blade obtained with AL and BEMT simulations are compared. Very good agreement is obtained for the velocity, while minor differences are found in the angle of attack, mainly in the inner part of the blade, i.e., $r/R < 0.4$. This difference is due to the hub effects that are considered in OpenFAST for the BEMT analysis but neglected in the AL approach. Near the hub, the aerodynamic load of the blade is decreased due to the flow blockage of the hub, which results in a reduction of the angle of attack. This is modeled in BEMT simulations through the root correction factor proposed by Prandtl [16]. In the AL-LES simulations, the hub and the tower are not included, and no correction is applied; therefore, the hub effect is not considered at all in the aerodynamic simulations. As a result, the angle of attack obtained with the BEMT approach is lower than the ones obtained with AL simulations, as observed in Fig. 4(b).

In general, root or hub loss corrections are not fundamental since the load contribution of the inner board to the overall rotor torque is insignificant compared to that of the outer board. The same occurs in noise production; the velocity in the outer board is so much higher than in the inner board that the noise is mainly coming from the outer part. Convergence analyses showed that the predicted far-field noise of the entire wind turbine does not vary more than 3 dB if only the outer 25% of the span is considered. The analyses are not included here for the sake of conciseness. This explains why the difference in the angle of attack for $r/R < 0.4$ between BEMT and AL results is not observed in the far-field noise prediction of the full rotor.

The results of the local angle of attack, apparent velocity, and Reynolds number along a single blade are averaged over one rotation, and the instantaneous results of all three blades are saved over one rotation period at 71 angular locations, with $\Delta t = 0.05$ s, which corresponds to a $\Delta \Psi = 5.1°$. For both cases, the initial transient effects are neglected.

### 4.3. Far-field noise prediction setup

In this section, the influence of the blade and the azimuth discretizations and the inputs model for Amiet's theory on the wind turbine noise prediction are addressed.

The noise prediction (including the XFOIL simulations for obtaining the boundary layer parameters) is conducted using a single processor calculation. The computational time was 1.01 CPU hours for the case of average AL-LES results and 71 CPU hours for the case of instantaneous AL-LES results. Note that the noise prediction code is not optimized for computational efficiency.

#### 4.3.1. Convergence analyses

The blade is divided into five segments of $AR \geq 3$, besides the tip. As shown in Fig. 5, the inner part of the blade that consists of cylinders ($r/R <= 0.2$) is neglected for the noise calculation. Fig. 6(a) shows the predicted wind turbine noise spectra for several blade divisions, where the number of divisions is obtained for a fixed $AR$. The blade division hardly affects the noise prediction, where the waves predicted by Amiet's theory for a 2D airfoil are more visible for the coarser case ($AR = 4$).

Fig. 6(b) shows the influence of the number of divisions of the azimuth angle. Less than 0.5 dB difference is observed in the maximum far-field noise power spectral density ($L_p$) when it is integrated for $n_\Psi = 10$ points and 70 points.

The prediction of the wind turbine noise is conducted with $n_\Psi = 20$, and $AR = 3$ since this is the finest discretization that can be obtained with a valid far-field noise approximation for Amiet's theory [60].

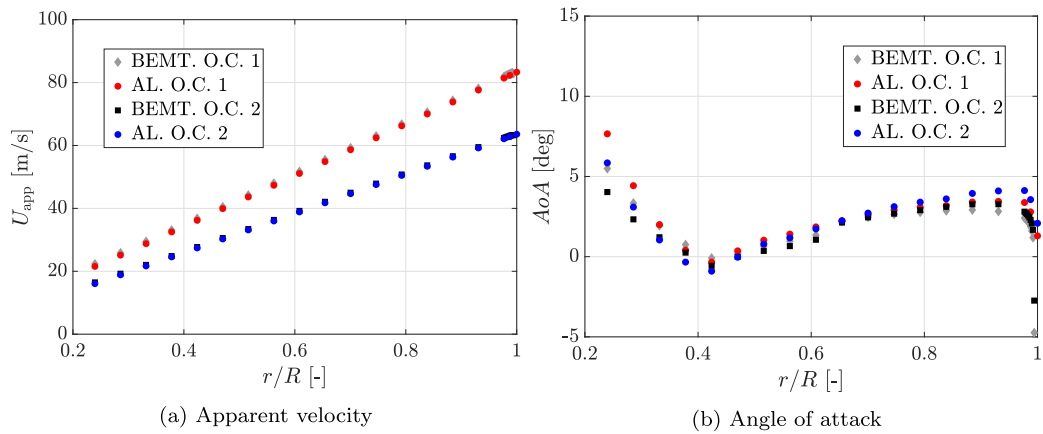

(a) Apparent velocity

(b) Angle of attack

**Fig. 4.** Comparison of apparent velocity and angle of attack distribution along the blade between actuator line averaged over one rotation and BEMT simulations.

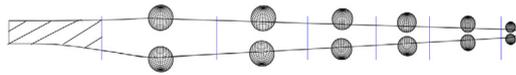

**Fig. 5.** Blade discretization for noise prediction.

*4.3.2. Influence of the inputs parameters for Amiet's theory*

Several models can be used to obtain the necessary input for Amiet's model to predict LE and TE noise. Fig. 7 shows the noise prediction using different models for calculating the turbulence spectrum and wall-pressure spectrum. Note that for a different turbulence spectrum (Fig. 7(a)) only the far-field noise in the low-frequency range is affected, which is dominated by the LE noise, whereas when the wall-pressure spectrum is varied, the mainly effect is observed in the high-frequency range where the TE noise dominates.

The von Kármán and Liepman models are considered references for leading-edge noise prediction [61] and for characterization of the turbulence spectrum for several applications, among which there are atmospheric flows [44]. The differences in the prediction using these two models (see Fig. 7(a)) are due to the theoretical nature of the von Kármán model for isotropic flows in the inertial subrange, and the semi-empirical nature of Liepmann's model [62]. On the other hand, the large differences observed for the turbulence distortion theory (TUD in the figure) compared to the other models are because this model was developed for cases where the integral length scale is much smaller than the airfoil chord to consider the distortion of the turbulence as it approaches the airfoil [43]. As the integral length scale for the case of a wind turbine is several times the order of magnitude of the airfoil chord, the turbulence distortion spectrum model highly underpredicts leading-edge noise.

There is no common agreement on the most accurate method for calculating the wall-pressure spectrum. Fig. 7(b) shows the wind turbine noise prediction using several wall-pressure spectrum models. The main difference among the wall-pressure spectrum models is that Goody's and Kamruzamann's models are semi-empirical models, whereas the TNO-Blake model is a semi-analytical model. Goody's model [52] is based on the scaling of the wall-pressure spectrum with the small and large turbulent scales within the boundary layer. This model results from measurements of a flat plate with no pressure gradient. Kamruzzamann's model [50] is based on Goody's model, but it considers measurements of different airfoils to take into account the effect of the adverse pressure gradient. TNO-Blake model [63,64] results from resolving analytical equations, e.g., the Poisson equation, to calculate the wall-pressure fluctuations. The solution considers velocities and turbulent profiles across the boundary layer. The biggest difference in the far-field noise spectra is observed between Goody's model and the TNO-Blake model. This difference is significantly reduced to a maximum difference of 2 dB if Goody's model is used only on the pressure side (PS) and the TNO-Blake model on the suction side due to the larger relative contribution of the suction side to the total noise. The difference between the case of the Kamruzzaman model and TNO-Blake model is 2 dB in the entire frequency range, with no relevant changes in the noise spectrum where the trailing-edge noise is dominant The reason why the Kamruzzamann model is closer to the TNO-Blake model than the Goody's model is because of the consideration of the pressure gradient that is not negligible for thick non-symmetric airfoils at relatively high angles of attack such as the ones used in wind turbines. It is worth mentioning that other empirical models can be used. However, they are not stable for every airfoil at the conditions along the blade since the input may be very distant from the ones the models are based on, resulting in unplausible values of far-field noise. The TNO-Blake model is used for the subsequent analyses because of its physical-based background.

*4.3.3. Noise source and inflow turbulence characteristics*

The dominant noise source of the wind turbine highly depends on the inflow turbulence and the operational conditions. Fig. 8(a) shows the wind turbine total noise and the contribution of the LE and TE noise for the O.C. 1 conditions of the wind turbine. LE noise is dominant for frequencies lower than 300 Hz, and TE noise is dominant for higher frequencies. Nonetheless, when analyzing the Overall A-Weighted Sound Pressure Level (OASPL), the trailing-edge noise is dominant over leading-edge noise along the blade section. This is because the trailing edge generates noise in a wider frequency range (see Fig. 9(b)).

The characteristics of the inflow turbulence significantly affect the LE noise. Thus, the dominant source of noise of a wind turbine varies with the inflow conditions. Fig. 10 shows the influence of the turbulence characteristics in the noise source. Note that the analyses were conducted, neglecting the effect of turbulence on the aerodynamic performance. Also, inflow turbulence is not considered in the AL simulations. In Fig. 10(a), the turbulence level is varied at a fixed integral length scale ($L = 300$ m). When the turbulence level is varied, LE noise is shifted in the spectral level without a change in the turbulence spectrum. Therefore, the frequency where LE noise is dominant is shifted to higher frequencies for higher inflow turbulence, and the total noise of the wind turbine is increased.

Fig. 10(b) shows the effect of the length scale in the LE noise for constant turbulence intensity ($Tu = 10.7\%$). The length scale was reduced from a typical value for atmospheric boundary layers ($L \approx 500$ m) to an implausible value smaller than the blade chord at the tip ($\approx 0.25$ m). A larger integral length scale shifts the energy content of the turbulence spectrum to lower frequencies, generating the $-5/3$ level decay as a function of the frequency typical of the turbulence spectrum inertial subrange at a much lower frequency. This results in a reduction of LE noise in the frequency range where the wind turbine noise is relevant for human hearing for higher length scales. At very low frequencies ($f < 200$ Hz), the tendency is the opposite, and larger length scales produce more noise since this is generated by the turbulence eddies that contain more energy directly related to the length scale.

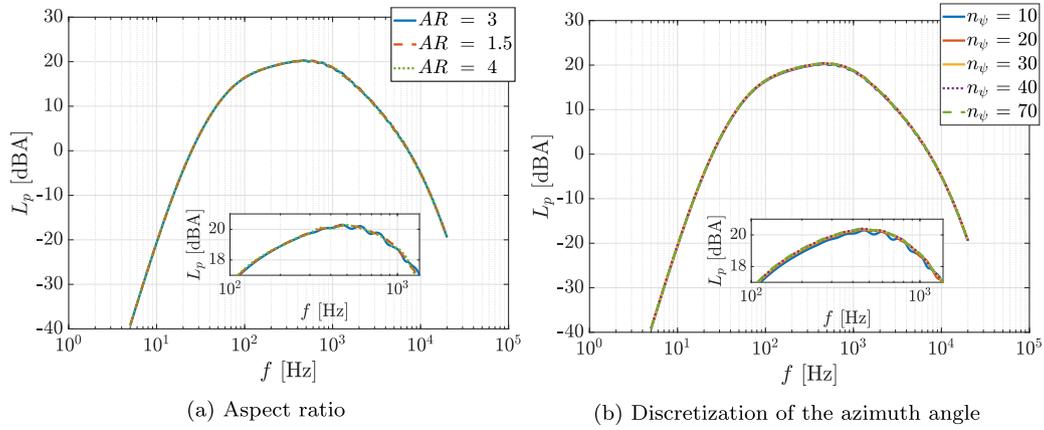

**Fig. 6.** Predicted wind turbine far-field noise for O.C. 1 for several aspect ratio and azimuth angle discretization.

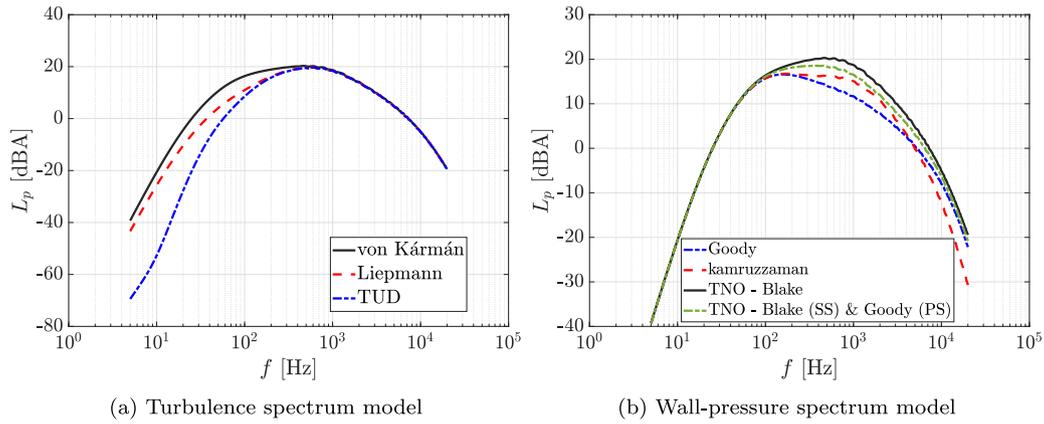

**Fig. 7.** Predicted LE and TE noise using different inflow turbulence (a) and wall-pressure spectrum (b) models for O.C. 1. TUD refers to the turbulence distortion model O.C. 1.

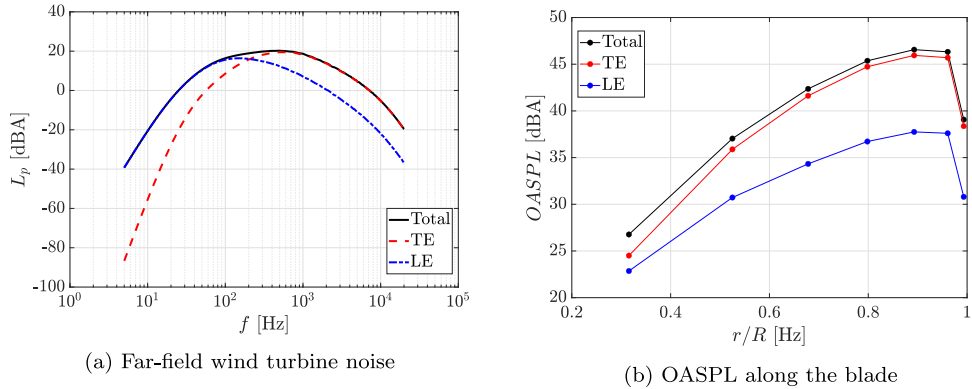

**Fig. 8.** Predicted LE and TE noise spectrum (a) and contribution to the integrated wind turbine noise (OASPL) along the blade (b). O.C. 1.

*4.4. Noise comparisons*

Fig. 11 shows the comparison of the predicted wind turbine noise with the measurements reported in Leloudas [58] for the two operational conditions shown in Table 1. For both operational conditions, the results include the predicted noise with the Amiet–Schlinker model discussed in this paper, the same prediction accounting for the atmospheric absorption, explained in Appendix, and the results obtained by Bresciani et al. [37], who synthesized the wind turbine noise model using the Harmonoise model and used BEMT to obtain blade loading. Another key difference with Ref. [37] is that they use RANS numerical simulations conducted in the commercial software STAR-CCM+, which, in principle, increases the fidelity of the simulations for the extraction of the boundary layer parameters but requires a more complex setup and a higher computational cost compared with XFOIL. For O. C. 1, an additional curve considering the instantaneous results of the actuator line simulations is included. In this case, the noise is calculated for each segment and blade at each azimuth location, considering the instantaneous results from AL, instead of the average values over one rotation.

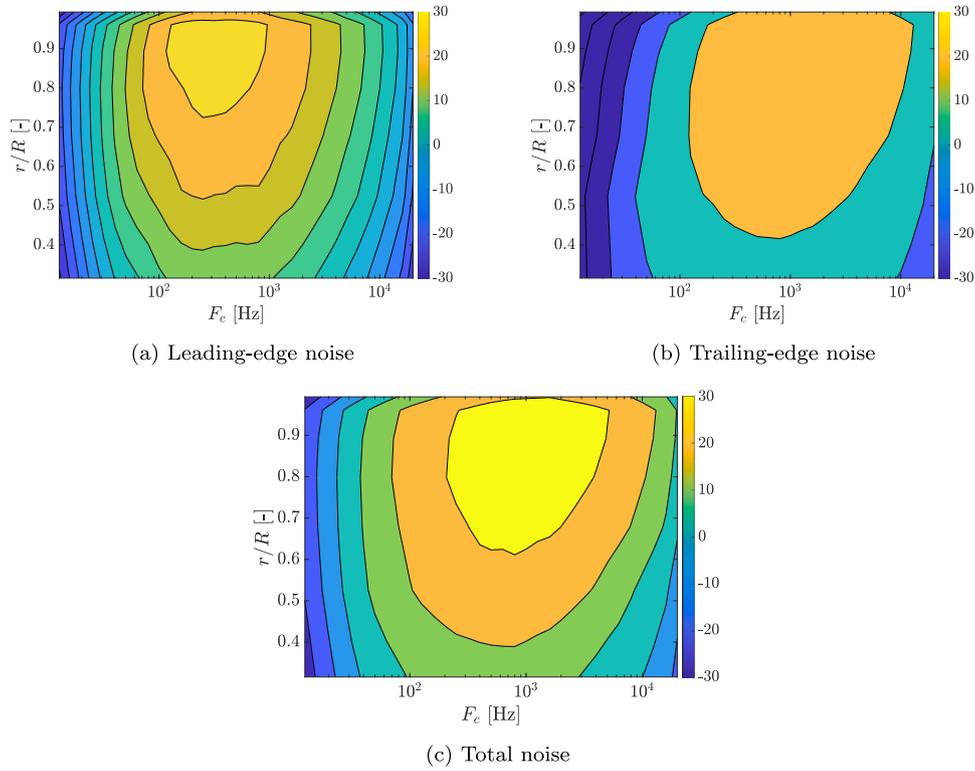

Fig. 9. Predicted $L_p$ [dBA] in one-third octave contours along the blade ($r/R$).

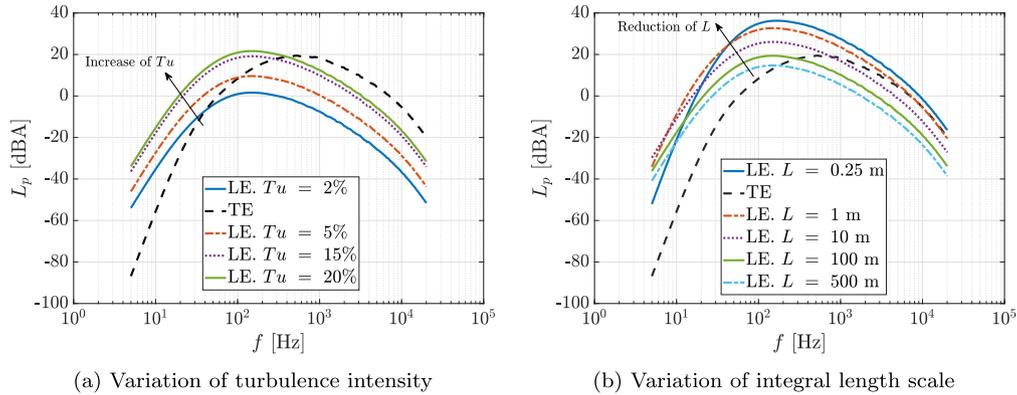

Fig. 10. Predicted wind turbine far-field noise for different turbulent inflow conditions for O.C. 1.

For both operational conditions, the wind turbine noise prediction agrees well with field measurements up to $f \approx 1$ kHz. For higher frequencies, the far-field noise is overpredicted, which is significantly improved when atmospheric absorption is taken into account. For this case, the prediction method presents a very good agreement with the measurements in the entire frequency range up to $f = 10$ kHz. There is a difference of about 4 dB in the frequency range between 20 and 100 Hz, which may be attributed to many factors such a slightly different turbulence level between the atmospheric flow and the simulations since it was obtained for a different operational condition.

This good agreement provides important validation to the methodology presented in this research. Furthermore, the wind turbine noise predicted with the proposed methodology matches the results of Bresciani et al. [37], which implies that the results obtained from XFOIL are in relatively good agreement with the ones obtained from the RANS simulations, with the main advantage of the proposed approach requiring no RANS simulations.

### 4.5. Instantaneous analyses

One of the main advantages of having the instantaneous results of the blade loading over a blade rotation is that variations on this scale can be accounted for in the aerodynamic analysis of the wind turbine and later in the wind turbine noise prediction without the need for computationally expensive noise simulations. Fig. 11(a) shows an excellent agreement between the far-field noise predicted using instantaneous and averaged results from AL simulations. In this case, the average values are representative of the variation along the rotor plane, and there is no sudden variation caused by transient effects of asymmetries on the rotor plane.

Fig. 12(a) shows the OASPL produced by each blade at every azimuth location and Fig. 12(b) depicts the OASPL for each blade as a function of the same relative azimuth position ($\Psi_{rot}$) over one rotation. The changes in the noise along the rotation for every blade, are due to the change of the noise directivity along the rotation with respect to a fixed observer. This plot also shows that at every azimuth location,

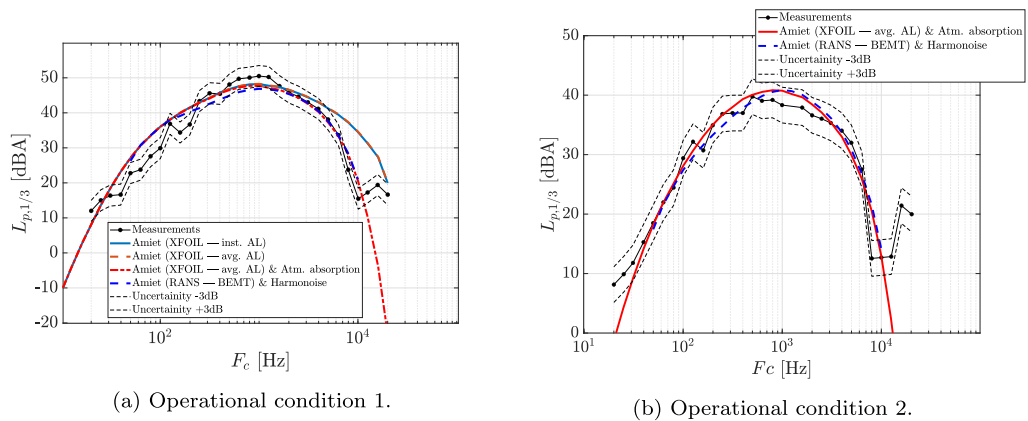

(a) Operational condition 1.

(b) Operational condition 2.

**Fig. 11.** Predicted and measured wind turbine far-field noise for several operational conditions shown in Table 1.

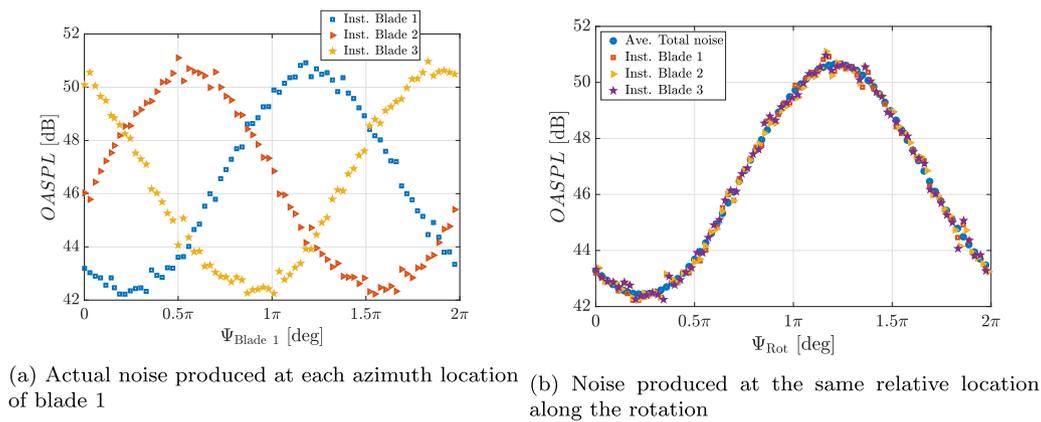

(a) Actual noise produced at each azimuth location of blade 1

(b) Noise produced at the same relative location along the rotation

**Fig. 12.** OASPL of each blade at each azimuth location for O.C. 1.

one blade is dominant over others. This is evident when analyzing Fig. 12(b). There is almost no deviation of the total wind turbine noise from the noise produced by each blade at the same relative position, demonstrating that there is no change in the blade loading at a specific azimuth location. In the situation where there is a blade loading change in a portion of the rotation, a difference in the noise produced among the blades at the same relative azimuth location would be observed.

Fig. 12(a) demonstrates that the differences in the noise production are mostly caused by the change of the directivity along the rotor location and not because of a change of the noise source.

Although in this research there is no consideration of non-asymmetries in the rotor plane or significant changes in the blade loading over a rotation, the presented methodology has the potential to consider more complex scenarios while remaining a low-cost method.

## 5. Conclusions

A method for coupling actuator line simulations with a low-cost fast-turnaround method for predicting wind turbine noise is presented. Leading- and trailing-edge noise are considered unique noise sources and are modeled by Amiet's theory. Several models are used to calculate the turbulence spectrum and the wall-pressure spectrum are discussed. Boundary layer information is taken from 2D data (computed with XFOIL), which makes this methodology a low-cost and highly efficient method for wind turbine noise prediction.

The actuator line simulations are validated by comparing the predicted thrust with measurements of the full scale wind turbine and by comparing the blade loading with BEMT simulations for several operational conditions. The wind turbine noise predictions match well with experimental measurements and predictions using other codes and more complex methodologies, which validated our approach.

A vital advantage of this method is that the wind turbine noise can be calculated using the blade loading at each azimuth location instead of averaged values over a rotation, allowing taking into account the effects of changes in the blade loading caused by, for example, gusts or wake of upstream wind turbines.

Finally, in this paper, the methodology has been validated for wind turbines, but it can be easily adapted to other rotating machinery, such as fans, helicopters, or rotors of unmanned aerial vehicles.

**CRediT authorship contribution statement**

**Laura Botero-Bolívar:** Writing – original draft, Visualization, Validation, Software, Methodology, Investigation, Formal analysis, Data curation, Conceptualization. **Oscar A. Marino:** Writing – original draft, Visualization, Validation, Software, Methodology, Investigation, Formal analysis, Conceptualization. **Cornelis H. Venner:** Writing – review & editing, Supervision, Resources, Project administration, Funding acquisition. **Leandro D. de Santana:** Writing – review & editing, Supervision, Resources, Project administration, Funding acquisition. **Esteban Ferrer:** Writing – review & editing, Validation, Resources, Methodology, Funding acquisition, Conceptualization.

**Declaration of competing interest**

The authors declare the following financial interests/personal relationships which may be considered as potential competing interests: Laura Botero Bolivar reports financial support was provided by University of Twente. Oscar Marino reports financial support was provided by Polytechnic University of Madrid. If there are other authors, they declare that they have no known competing financial interests or personal relationships that could have appeared to influence the work reported in this paper.

**Data availability**

The data that support the findings of this study are available from the corresponding author, Botero-Bolívar, L., upon reasonable request. Furthermore, the algorithm to predict wind turbine noise is available in Botero-Bolívar (2023) [54].

**Acknowledgments**

The authors would like to express their gratitude for the support of the Agencia Estatal de Investigación for the grant "Europa Excelencia" for the project EUR2022-134041 funded by MCIN/AEI/10.13039/501100011033 and the European Union NextGenerationEU/PRTR. The author thankfully acknowledges the computer resources at MareNostrum and the technical support provided by Barcelona Supercomputing Center (RES-IM-2022-3-0023).

This research has received funding from the European Commission through the H2020-MSCA-ITN-209 project zEPHYR (grant agreement No 860101) and the European Union (ERC, Off-coustics, project number 101086075). Views and opinions expressed here are those of the author(s) only and do not necessarily reflect those of the European Union or the European Research Council. Neither the European Union nor the granting authority can be held responsible for them.

**Appendix. Atmospheric attenuation**

The atmospheric absorption is applied after the calculation of the segment noise since it depends on the distance between the noise source and the observer as:

$$A_{\text{atm}} = \alpha_a d_n; \tag{A.1}$$

where $d_n$ is the distance from the noise source to the observer and $\alpha_a$ is the attenuation in dB/m of travel through the atmosphere for a pure tone frequency given by the standard ANSI/ASA S1.26 [5,65]. The atmospheric conditions used in this research for the calculation of the atmospheric attenuation are: $T = 15$ °C and $T_{\text{ref}} = 20$ °C; $P = 98$ kPa and $P_{\text{ref}} = 101.325$ kPa; and $h = 86\%$; where $T$ and $T_{\text{ref}}$ are the source and reference temperatures, $P$ and $P_{\text{ref}}$ are the source and reference atmospheric pressures, and $h$ is the relative humidity.

**References**

[1] J.L. Davy, K. Burgemeister, D. Hillman, Wind turbine sound limits: Current status and recommendations based on mitigating noise annoyance, Appl. Acoust. 140 (2018) 288–295, http://dx.doi.org/10.1016/j.apacoust.2018.06.009, URL https://www.sciencedirect.com/science/article/pii/S0003682X17311428.

[2] S. Oerlemans, Wind turbine noise: primary noise sources, Technical Report, National Aerospace Laboratory NLR, 2011.

[3] S. Oerlemans, Reduction of wind turbine noise using blade trailing edge devices, in: 22$^{nd}$ AIAA/CEAS Aeroacoustics Conference, American Institute of Aeronautics and Astronautics, 2016, p. 3018, http://dx.doi.org/10.2514/6.2016-3018.

[4] S. Buck, S. Oerlemans, S. Palo, Experimental characterization of turbulent inflow noise on a full-scale wind turbine, J. Sound Vib. 385 (2016) 219–238, http://dx.doi.org/10.1016/j.jsv.2016.09.010, URL https://www.sciencedirect.com/science/article/pii/S0022460X16304564.

[5] C.H. Hansen, C.J. Doolan, K.L. Hansen, Wind farm noise: measurement, assessment, and control, John Wiley & Sons, 2017.

[6] J.E. Ffowcs Williams, D.L. Hawkings, Sound generation by turbulence and surfaces in arbitrary motion, Philos. Trans. R. Soc. Lond. Series A 264 (1151) (1969) 321–342.

[7] F. Farassat, G. Succi, A review of propeller discrete frequency noise prediction technology with emphasis on two current methods for time domain calculations, J. Sound Vib. 71 (3) (1980) 399–419, http://dx.doi.org/10.1016/0022-460x(80)90422-8.

[8] P. di Francescantonio, A new boundary integral formulation for the prediction of sound radiation, J. Sound Vib. 202 (4) (1997) 491–509, http://dx.doi.org/10.1006/jsvi.1996.0843.

[9] D.P. Lockard, An efficient, two-dimensional implementation of the ffowcs williams and hawkings equation, J. Sound Vib. 229 (4) (2000) 897–911.

[10] D. Lockard, A comparison of ffowcs williams-hawkings solvers for airframe noise applications, in: 8th AIAA/CEAS Aeroacoustics Conference & Exhibit, American Institute of Aeronautics and Astronautics, 2002, http://dx.doi.org/10.2514/6.2002-2580.

[11] G. Ghorbaniasl, Z. Huang, L. Siozos-Rousoulis, C. Lacor, Analytical acoustic pressure gradient prediction for moving medium problems, Proc. R. Soc. A 471 (2184) (2015) 20150342, http://dx.doi.org/10.1098/rspa.2015.0342.

[12] A. Tadamasa, M. Zangeneh, Numerical prediction of wind turbine noise, Renew. Energy 36 (7) (2011) 1902–1912, http://dx.doi.org/10.1016/j.renene.2010.11.036.

[13] R. Schlinker, R. Amiet, Helicopter rotor trailing edge noise, in: 7th Aeroacoustics Conference, 1981, p. 2001.

[14] Y. Rozenberg, M. Roger, S. Moreau, Rotating blade trailing-edge noise: Experimental validation of analytical model, AIAA J. 48 (5) (2010) 951–962, http://dx.doi.org/10.2514/1.43840.

[15] Y. Tian, B. Cotté, Wind turbine noise modeling based on amiet's theory: Effects of wind shear and atmospheric turbulence, Acta Acust. United Acust. 102 (4) (2016) 626–639, http://dx.doi.org/10.3813/AAA.918979.

[16] P. Bortolotti, M. Masciola, S. Ananthan, M.J. Schmidt, J. Rood, N. Mendoza, M. Hai, A. Sharma, K. Shaler, K. Bendl, P. Schuenemann, P. Sakievich, E. Quon, M.R. Phillips, N. Kusouno, A. Gonzalez Salcedo, T. Martinez, R. Corniglion, Openfast, 2022, http://dx.doi.org/10.5281/zenodo.6324288, URL https://github.com/OpenFAST/openfast/tree/v3.1.0.

[17] J. Colas, A. Emmanuelli, D. Dragna, P. Blanc-Benon, B. Cotté, R.J.A.M. Stevens, Wind turbine sound propagation: Comparison of a linearized Euler equations model with parabolic equation methods, J. Acoust. Soc. Am. 154 (3) (2023) 1413–1426, http://dx.doi.org/10.1121/10.0020834.

[18] S. Sinayoko, M. Kingan, A. Agarwal, Trailing edge noise theory for rotating blades in uniform flow, Proc. R. Soc. A 469 (2157) (2013) 20130065.

[19] T.F. Brooks, D. Pope, M.A. Marcolini, Airfoil self-noise and prediction, Technical Report, NASA Reference Publication 1218, 1989.

[20] R. Amiet, Noise due to turbulent flow past a trailing edge, J. Sound Vib. 47 (3) (1976) 387–393, http://dx.doi.org/10.1016/0022-460X(76)90948-2.

[21] R. Amiet, Acoustic radiation from an airfoil in a turbulent stream, J. Sound Vib. 41 (4) (1975) 407–420, http://dx.doi.org/10.1016/S0022-460X(75)80105-2.

[22] J.N. Sorensen, W.Z. Shen, Numerical modeling of wind turbine wakes, J. Fluids Eng. 124 (2) (2002) 393–399, http://dx.doi.org/10.1115/1.1471361.

[23] L. Liu, L. Franceschini, D.F. Oliveira, F.C. Galeazzo, B.S. Carmo, R.J.A.M. Stevens, Evaluating the accuracy of the actuator line model against blade element momentum theory in uniform inflow, Wind Energy 25 (6) (2022) 1046–1059, http://dx.doi.org/10.1002/we.2714, arXiv:https://onlinelibrary.wiley.com/doi/pdf/10.1002/we.2714.

[24] J.N. Sørensen, R.F. Mikkelsen, D.S. Henningson, S. Ivanell, S. Sarmast, S.J. Andersen, Simulation of wind turbine wakes using the actuator line technique, Phil. Trans. R. Soc. A 373 (2035) (2015) 20140071, http://dx.doi.org/10.1098/rsta.2014.0071, arXiv:https://royalsocietypublishing.org/doi/pdf/10.1098/rsta.2014.0071.

[25] H. Debertshäuser, W. Shen, W. Zhu, Development of a high-fidelity noise prediction and propagation model for noise generated from wind turbines, in: Proceedings. 6th International Meeting on Wind Turbine Noise, 2015, 6th International Meeting on Wind Turbine Noise ; Conference date: 20-04-2015 Through 23-04-2015.

[26] R. Ewert, W. Schröder, Acoustic perturbation equations based on flow decomposition via source filtering, J. Comput. Phys. 188 (2) (2003) 365–398.

[27] H. Debertshaeuser, W.Z. Shen, W.J. Zhu, Aeroacoustic calculations of wind turbine noise with the actuator line/ Navier-Stokes technique, in: 34th Wind Energy Symposium, American Institute of Aeronautics and Astronautics, 2016, http://dx.doi.org/10.2514/6.2016-0750.

[28] W.J. Zhu, W.Z. Shen, E. Barlas, F. Bertagnolio, J.N. Sørensen, Wind turbine noise generation and propagation modeling at DTU wind energy: A review, Renew. Sustain. Energy Rev. 88 (2018) 133–150, http://dx.doi.org/10.1016/j.rser.2018.02.029.

[29] Z. Cheng, F.-S. Lien, E. Yee, H. Meng, A unified framework for aeroacoustics simulation of wind turbines, Renew. Energy 188 (2022) 299–319, http://dx.doi.org/10.1016/j.renene.2022.01.120.

[30] J. Schluntz, R.H. Willden, An actuator line method with novel blade flow field coupling based on potential flow equivalence, Wind Energy 18 (8) (2015) 1469–1485, http://dx.doi.org/10.1002/we.1770, arXiv:https://onlinelibrary.wiley.com/doi/pdf/10.1002/we.1770.

[31] L. Liu, L. Franceschini, D. Oliveira, F. Galeazzo, B. Carmo, R. Stevens, Evaluating the accuracy of the actuator line model against blade element momentum theory in uniform inflow, Wind Energy (2022) 1–14, http://dx.doi.org/10.1002/we.2714.

[32] N. Troldborg, J. Sørensen, R. Mikkelsen, Numerical simulations of wake characteristics of a wind turbine in uniform inflow, Wind Energy 13 (2010) 86–99, http://dx.doi.org/10.1002/we.345.

[33] L. Martínez-Tossas, M. Churchfield, C. Meneveau, Large eddy simulation of wind turbine wakes: detailed comparisons of two codes focusing on effects of numerics and subgrid modeling, J. Phys. Conf. Ser. 625 (2015) 12024, http://dx.doi.org/10.1088/1742-6596/625/1/012024.

[34] E. Ferrer, G. Rubio, G. Ntoukas, W. Laskowski, O.A. Marino, S. Colombo, A. Mateo-Gabín, H. Marbona, F. Manrique de Lara, D. Huergo, J. Manzanero, A.M. Rueda-Ramírez, D.A. Kopriva, E. Valero, HORSES3D: A high-order discontinuous Galerkin solver for flow simulations and multi-physics applications, Comput. Phys. Comm. 287 (2023) 108700, http://dx.doi.org/10.1016/j.cpc.2023.108700, URL https://github.com/loganoz/horses3d.

[35] E. Ferrer, G. Rubio, G. Ntoukas, W. Laskowski, O. Marino, S. Colombo, A. Mateo-Gabín, H. Marbona, F. Manrique de Lara, D. Huergo, J. Manzanero, A. Rueda-Ramírez, D. Kopriva, E. Valero, HORSES3D: A high-order discontinuous Galerkin solver for flow simulations and multi-physics applications, Comput. Phys. Comm. 287 (2023) 108700, http://dx.doi.org/10.1016/j.cpc.2023.108700, URL https://www.sciencedirect.com/science/article/pii/S0010465523000450.

[36] O.A. Marino, E. Ferrer, E. Valero, O. Ferret, Aeroacoustic simulations of 3D airfoil sections using a high order discontinuous Galerkin solver, in: AIAA SCITECH 2022 Forum, http://dx.doi.org/10.2514/6.2022-0413, arXiv:https://arc.aiaa.org/doi/pdf/10.2514/6.2022-0413.

[37] A.C. Bresciani, J. Maillard, S. Le Bras, L.D. de Santana, Wind turbine noise synthesis from numerical simulations, in: AIAA AVIATION 2023 Forum, 2023, p. 3643, http://dx.doi.org/10.2514/6.2023-3643.

[38] M. Roger, On broadband jet–ring interaction noise and aerofoil turbulence-interaction noise, J. Fluid Mech. 653 (2010) 337–364, http://dx.doi.org/10.1017/S0022112010000285.

[39] M. Roger, S. Moreau, Back-scattering correction and further extensions of amiet's trailing-edge noise model. Part 1: theory, J. Sound Vibrat. 286 (3) (2005) 477–506, http://dx.doi.org/10.1016/j.jsv.2004.10.054.

[40] A.P. Bresciani, S. Le Bras, L.D. de Santana, Generalization of amiet's theory for small reduced-frequency and nearly-critical gusts, J. Sound Vib. 524 (2022) 116742, http://dx.doi.org/10.1016/j.jsv.2021.116742.

[41] T. von Kármán, Progress in the statistical theory of turbulence, Proc. Natl. Acad. Sci. 34 (11) (1948) 530–539, http://dx.doi.org/10.1073/pnas.34.11.530.

[42] H.W. Liepmann, J. Laufer, Investigations of free turbulent mixing, Technical Report NACA-TN-1257, NASA, 1947.

[43] L.D. Santana, J. Christophe, C. Schram, W. Desmet, A rapid distortion theory modified turbulence spectra for semi-analytical airfoil noise prediction, J. Sound Vib. 383 (2016) 349–363, http://dx.doi.org/10.1016/j.jsv.2016.07.026, URL https://www.sciencedirect.com/science/article/pii/S0022460X1630356X.

[44] M. Emes, M. Arjomandi, R. Kelso, F. Ghanadi, Integral length scales in a low-roughness atmospheric boundary layer, AWES, 2016.

[45] R.B. Stull, An introduction to boundary layer meteorology, vol. 13, Springer Science & Business Media, 2012.

[46] B. Kale, ABL simulations with uncertain weather parameters and impact on WT performance and near-field noise (Ph.D. thesis), Ph.D. thesis, Universidad Politécnica de Madrid, 2023.

[47] R. Amiet, Noise due to turbulent flow past a trailing edge, J. Sound Vib. 47 (3) (1976) 387–393, http://dx.doi.org/10.1016/0022-460X(76)90948-2.

[48] G.M. Corcos, The structure of the turbulent pressure field in boundary-layer flows, J. Fluid Mech. 18 (3) (1964) 353–378, http://dx.doi.org/10.1017/S002211206400026X.

[49] O. Stalnov, P. Chaitanya, P.F. Joseph, Towards a non-empirical trailing edge noise prediction model, J. Sound Vib. 372 (2016) 50–68, http://dx.doi.org/10.1016/j.jsv.2015.10.011.

[50] M. Kamruzzaman, D. Bekiropoulos, T. Lutz, W. Würz, E. Krämer, A semi-empirical surface pressure spectrum model for airfoil trailing-edge noise prediction, Int. J. Aeroacoust. 14 (5–6) (2015) 833–882, http://dx.doi.org/10.1260/1475-472X.14.5-6.833.

[51] S. Lee, L. Ayton, F. Bertagnolio, S. Moreau, T.P. Chong, P. Joseph, Turbulent boundary layer trailing-edge noise: Theory, computation, experiment, and application, Prog. Aerosp. Sci. 126 (2021) 100737, http://dx.doi.org/10.1016/j.paerosci.2021.100737, URL.

[52] M. Goody, Empirical spectral model of surface pressure fluctuations, AIAA J. 42 (9) (2004) 1788–1794, http://dx.doi.org/10.2514/1.9433.

[53] M. Drela, XFOIL: An analysis and design system for low Reynolds number airfoils, in: Low Reynolds Number Aerodynamics, Springer, 1989, pp. 1–12.

[54] L. Botero-Bolívar, Wind turbine noise prediction, 2023, http://dx.doi.org/10.5281/zenodo.8325974, URL https://github.com/lauraboterob/WindTurbine_Noise_prediction.

[55] B. Cotté, S. Roy, D. Raus, R. Oueini, Towards a semi-empirical trailing edge noise model valid for attached and separated turbulent boundary layers, in: 28th AIAA/CEAS Aeroacoustics 2022 Conference, 2022, p. 3103, http://dx.doi.org/10.2514/6.2022-3103.

[56] F. Bertagnolio, H.A. Madsen, A. Fischer, C. Bak, A semi-empirical airfoil stall noise model based on surface pressure measurements, J. Sound Vib. 387 (2017) 127–162, http://dx.doi.org/10.1016/j.jsv.2016.09.033.

[57] T. Suresh, L. Botero, O. Szulc, P.H. Flaszynski, Numerical investigation of acoustic effect of vortex generators on a wind turbine rotor blade, in: AIAA SCITECH 2024 Forum, 2024, p. 2102.

[58] G. Leloudas, Optimization of Wind Turbines with respect to Noise, Technical University of Denmark, Technical University of Denmark, Anker Engelunds Vej 1, Building 101A, 2800 Kgs. Lyngby, 2006, p. 66.

[59] J. Christophe, S. Buckingham, C. Schram, S. Oerlemans, zEPHYR - Large On Shore Wind Turbine Benchmark, Zenodo, 2022, http://dx.doi.org/10.5281/zenodo.7323750.

[60] M. Roger, S. Moreau, Extensions and limitations of analytical airfoil broadband noise models, Int. J. Aeroacoust. 9 (3) (2010) 273–305.

[61] D. Lewis, J. de Laborderie, M. Sanjosé, S. Moreau, M.C. Jacob, V. Masson, Parametric study on state-of-the-art analytical models for fan broadband interaction noise predictions, J. Sound Vib. 514 (2021) 116423, http://dx.doi.org/10.1016/j.jsv.2021.116423, URL https://www.sciencedirect.com/science/article/pii/S0022460X21004673.

[62] S. Moreau, M. Roger, Competing broadband noise mechanisms in low-speed axial fans, AIAA J. 45 (1) (2007) 48–57, http://dx.doi.org/10.2514/1.14583.

[63] O. Stalnov, P. Chaitanya, F.J. Phillip, Towards a non-empirical trailing edge noise prediction model, J. Sound Vib. 372 (2016) 50–68, http://dx.doi.org/10.1016/j.jsv.2015.10.011.

[64] W.K. Blake, Mechanics of flow-induced sound and vibration, Volume 2: Complex flow-structure interactions, Academic Press, 2017.

[65] ANSI/ASA S1 26-2014, Methods for Calculation of the Absorption of Sound by the Atmosphere, American National Standards Institute Melville, NY, 2014.